\documentclass[journal]{IEEEtran}

\usepackage{subcaption}
\usepackage[pdftex]{graphicx}
\usepackage{amsmath}
\usepackage{algorithmic}
\usepackage{cleveref}
\usepackage{url}
\usepackage{graphicx, float}
\usepackage{caption,lipsum}

\begin{document}

\title{In situ visualization of regional-scale natural hazards with Galaxy and Material Point Method}

\author{Greg Abram$^1$, %~\IEEEmembership{Fellow,~OSA,}
        Andrew Solis$^1$,
        Yong Liang$^2$,
        and~Krishna Kumar$^{3}$% <-this % stops a space
\thanks{$^1$Texas Advanced Computing Center, University of Texas at Austin, TX,.}% <-this % stops a space
\thanks{$^2$University of California at Berkeley, CA,.}% <-this % stops a space
\thanks{$^{3}$Department of Civil, Architectural and Environmental Engineering, University of Texas at Austin, TX, 78712 USA e-mail: (krishnak@utexas.edu).}% <-this % stops a space
\thanks{Manuscript received September 3rd, 2021. Revised on January 31st, 2022.}}

\markboth{IEEE Computing in Science and Engineering (CiSE), Special Issue 2021}%
{}


%\maketitle

\maketitle

\begin{abstract}
Visualizing regional-scale landslides is essential to conveying the threat of natural hazards to stakeholders and policymakers. Traditional visualization techniques are restricted to post-processing a limited subset of simulation data and are not scalable to rendering regional-scale models. In situ visualization is a technique of rendering simulation data in real-time, i.e., rendering visuals in tandem while the simulation is running. This study develops a scalable N:M interface architecture to visualize regional-scale landslides. We demonstrate the scalability of the architecture by simulating the long runout of the 2014 Oso landslide using the Material Point Method coupled with the Galaxy ray tracing engine rendering 4.2 million material points as spheres. In situ visualization has an amortized runtime increase of 2\% compared to non-visualized simulations. The developed approach can achieve in situ visualization of regional-scale landslides with minimal impact on the simulation process.
\end{abstract}

%\begin{figure*}
%\includegraphics[width=\linewidth]{figs/teaser.png}
%\caption{Galaxy renderings of the Oso landslide early(L) and late(R) in the runout. The color gradient shows the amount of particle displacements from %their original position.}
%\label{Fig:runout}
%\end{figure*} 
 
 \begin{figure*}[!tbp]
    \begin{subfigure}[b]{0.333\textwidth}
        \includegraphics[width=\textwidth]{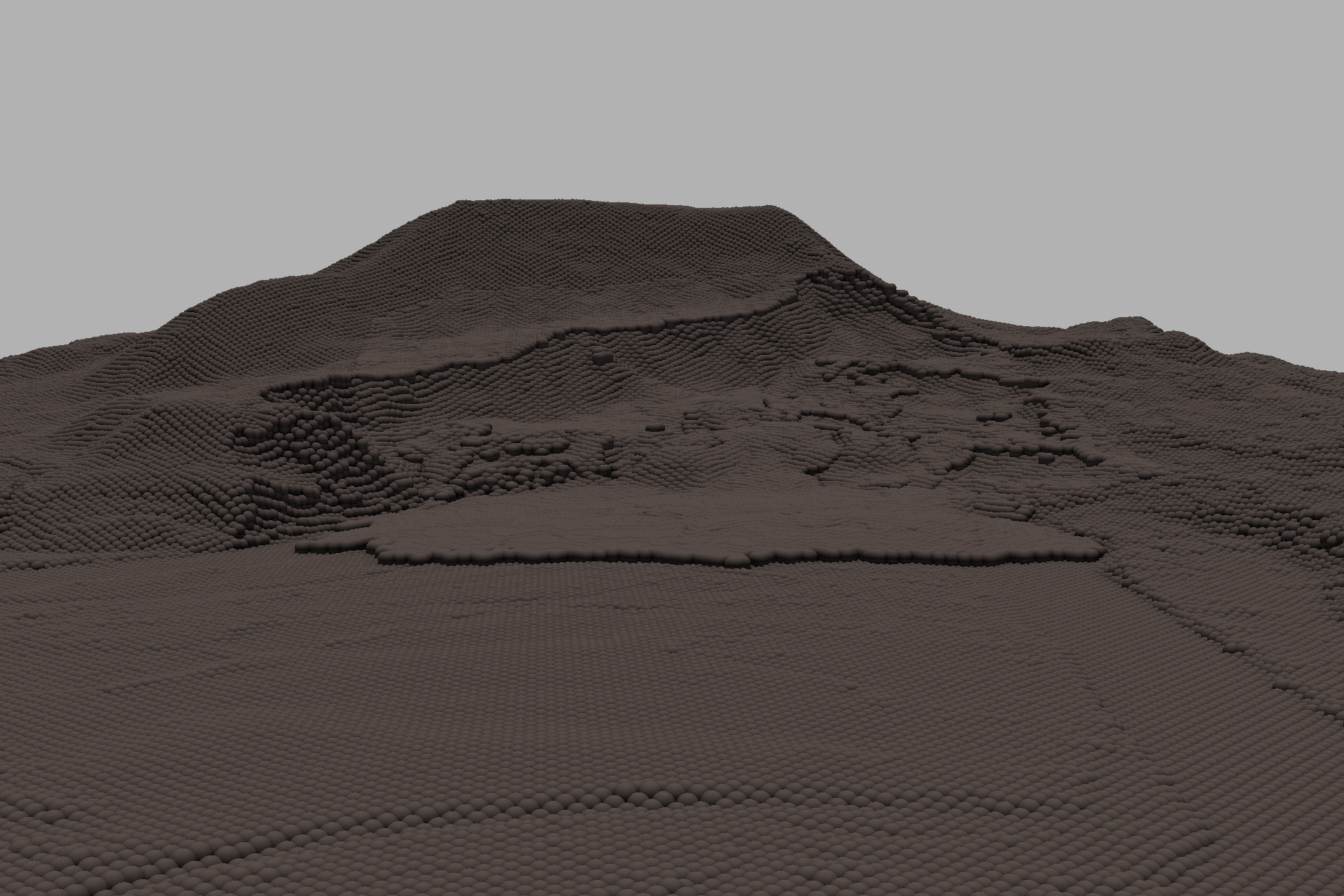}
        \caption{runout after 10 s}
        %\label{fig:oso-10} 
    \end{subfigure}
    \begin{subfigure}[b]{0.334\textwidth}
        \includegraphics[width=\textwidth]{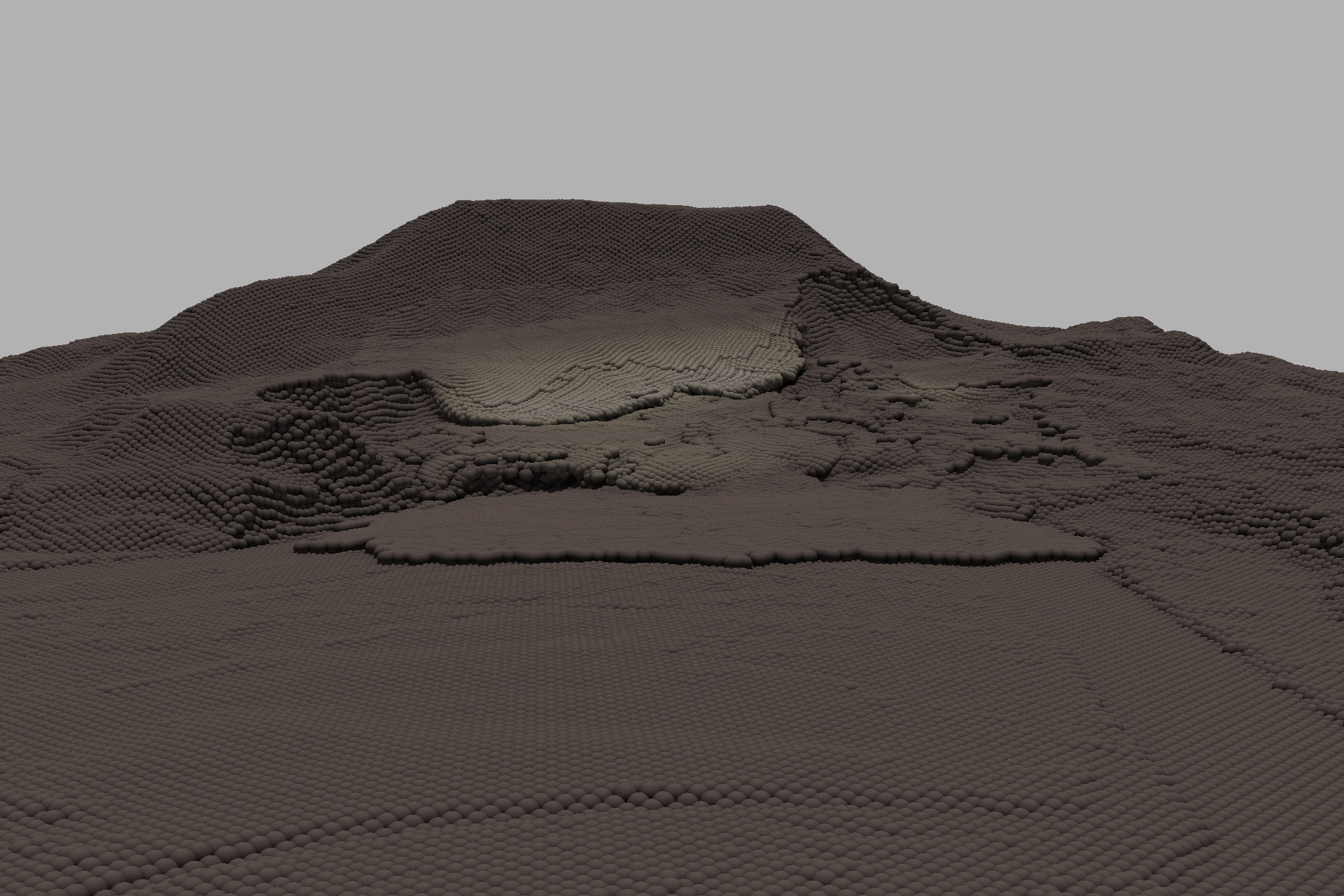}
        \caption{runout after 15 s}
        %\label{fig:oso-15} 
    \end{subfigure}
    \begin{subfigure}[b]{0.333\textwidth}
        \includegraphics[width=\textwidth]{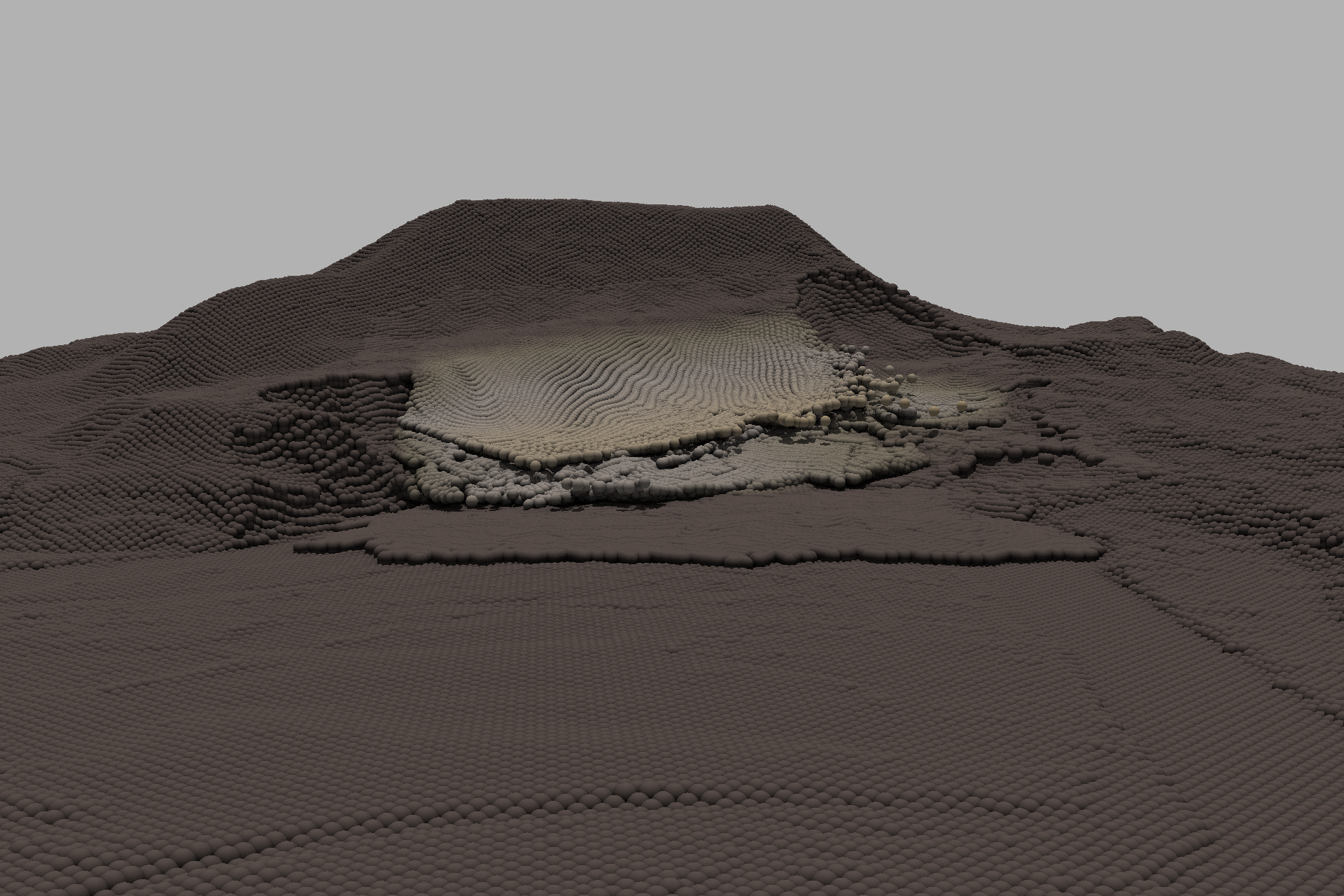}
        \caption{runout after 20 s}
    \end{subfigure}
    \caption{Early stages of the simulated landslide runout visualized using in situ rendering of CB-Geo MPM simulation with TACC Galaxy. The color gradient shows the amount of particle displacements from their original position.}
    \label{fig:oso3}
\end{figure*}


\IEEEpeerreviewmaketitle

\section{Overview and motivation}
% - why simulating earthquakes is important
% - why such sim is hard
% - brief description of MPM and benefits, cite to marker paper
% - why in situ vis is needed / benefits
% - Galaxy + MPM = awesomeness
\IEEEPARstart{L}{andslide} runouts are regional-scale events that can bury whole towns (e.g., 2018 Southern California mudflows that followed a series of wildfires, 2014 Oso landslide caused 43 fatalities in the US) or devastate entire regions (e.g., 2016 Kaikoura, New Zealand earthquake recorded 100,000 landslides within a 12,000 sq. km area). The frequency of these regional-scale landslides is increasing with devastating earthquakes, extreme precipitation, and wildfires caused by climate change. Even in these risk-prone communities, where the residents are aware of the threat posed by the landslides, the residents repeatedly show indifference to these threats and consequently fail to act. The human capacity to deny danger endangers lives and property~\cite{kim2020public}. In 2021, twenty extreme events cost the United States \$145 billion in damages. The underlying problem is the uncertainty of the actual event and a lack of understanding of its potential impacts. How to communicate the reality of potential risk to a diverse group of individuals, who must understand and support a complex set of actions that reduce the risk for the whole community?
  
Visualization is the key to communicating scientific results effectively to engineering decision-makers and the public. Given the limited cognitive capacity in humans, visualizing the complex threats as images enhances the brain's capacity to perceive opportunities and make decisions. Visual representation of complex information reduces the cognitive load on human information processing and increases human ``absorptive capacity" for problem-solving~\cite{kahneman1983cost}. However, these visualizations must be physically sound to be perceived as realistic and accepted as valid.  \Cref{fig:oso3} shows a realistic rendering of an early-stage landslide runout. Visualization is also context and user-specific, i.e., different users may be interested in different aspects of the experimental or simulation data sets. Hence, the same data set may be represented in several forms – perhaps at varying levels of detail, emphasizing or deemphasizing different regions and features, and employing different visualization techniques to best present the information.

\begin{figure}[tbp]
    \includegraphics[width=\linewidth]{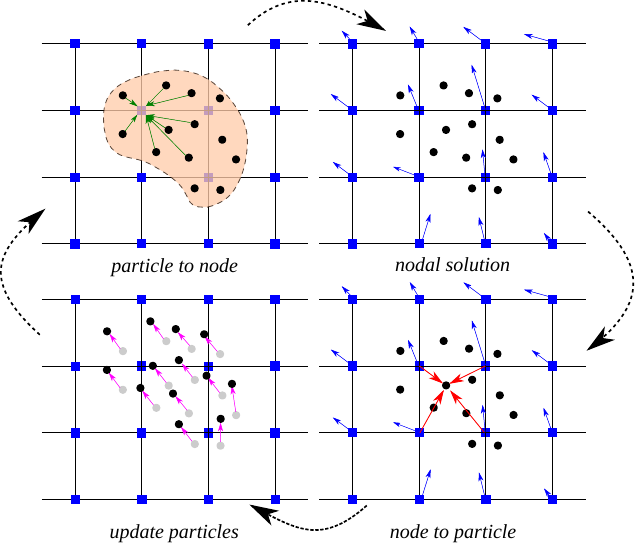}
   \caption{Illustration of the MPM algorithm (1) A representation of material points overlaid on a computational grid. Arrows represent material point state vectors (e.g., mass, volume, and velocity) is projected to the nodes of the computational grid. (2) The equations of motion are solved onto the nodes, resulting in updated nodal velocities and positions. (3) The updated nodal kinematics is interpolated back to the material points. (4) The state of the material points is updated, and the computational grid is reset.}
   \label{fig:mpm}
\end{figure}

Despite the landslide risks, most regional-scale landslide hazard analyses do not consider downstream impacts and the aerial extent of debris-flow runouts. Traditional numerical methods such as the Finite Element Method (FEM) primarily focus on the onset of failure but provide limited information on the post-failure runout mechanisms due to mesh distortions at large  deformations~\cite{soga2016trends}. Modeling the impacts of the potential runout of large landslides is possible with tools such as Material Point Method (MPM)~\cite{kumar2019scalable}. MPM is a mesh-free method that discretizes the domain as a collection of \textit{material points} moving on a background grid, and Newton's laws of motion determine their deformations. ~\Cref{fig:mpm} shows a typical MPM computation cycle. In this study, we employ the CB-Geo MPM code (\url{https://github.com/cb-geo/mpm}). For more information on MPM and the code implementation, refer to Kumar et. al. ~\cite{kumar2019scalable}. MPM simulations of the landslide runout can be translated into compelling visualizations to inform decision-makers of potential hazards (see~\cref{fig:oso3}).

\begin{figure}[!htbp]
    \centering
    \includegraphics[width=\linewidth]{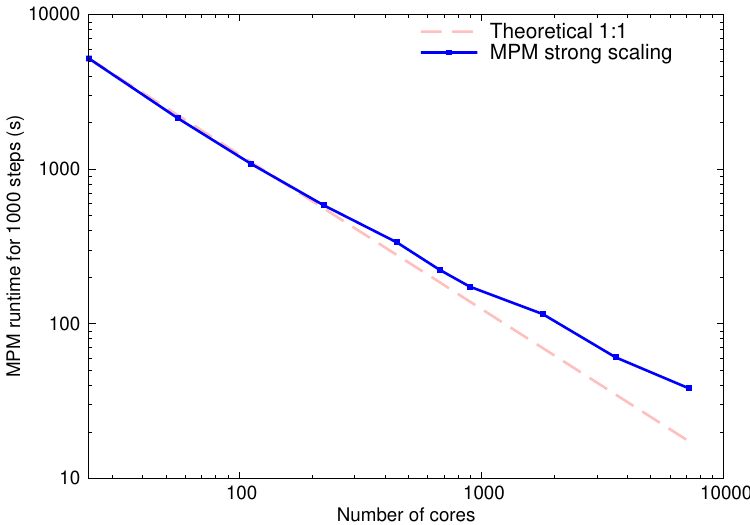}
    \caption{Strong scaling results of CB-Geo MPM simulating 7 million material points on TACC Stampede2.} 
    \label{fig:scaling}
\end{figure}

An MPM simulation of regional-scale landslide requires millions to billions of material points and background grid nodes. Scaling applications to model region-scale hazards requires  high-performance computing (HPC) architectures with efficient Message-Passing Interface (MPI). The CB-Geo MPM code exploits a hybrid MPI+X approach to achieve regional-scale simulation of landslides. In addition, as the landslide runout progresses, the workload distributed across the compute nodes becomes imbalanced. We adopt a distributed graph-partitioning technique to load balance both the background mesh and particle distribution dynamically.  \Cref{fig:scaling} shows the hybrid MPI+X approach of CB-Geo MPM offers linear scaling up to 10k cores for simulating 7 million material points on Texas Advanced Computing Center (TACC) Stampede2 with Intel Xeon Skylake nodes. 

Data retrieval and visualization in HPC applications have long been a bottleneck. Current techniques for visualizing regional-scale simulations involve writing a temporal slice of a subset of the data to disk, often at a much coarser resolution than the original data, which leads to a significant portion of information being disregarded and potentially lost. In practice, the simulation data are only stored at specific time steps, often every 1,000 or 10,000 steps, followed by post-processing of data. A  regional-scale simulation would output up to terabytes of data, resulting in a significant portion of the compute time spent on I/O~\cite{byna2020exahdf5}. Nevertheless, such regional-scale discrete geometry visualization will not be efficient or scalable without leveraging distributed computing. 

Running the visualization and simulation in tandem avoids data transfer bottleneck, enabling scientists to monitor and modify simulation parameters in real-time. In situ visualization libraries such as ParaView Catalyst and SENSEI offer the ability to couple existing data models (such as VTK) to query regions of interest and to render simulations. However, scaling rendering engines to regional-scale simulations remains a challenge requiring billions of lightpaths to be rendered. Most rendering engines only sample a limited number of random light paths at each step. Thus, the visual quality of rendered images suffers from estimator variance, which appears as visually distracting noise~\cite{yang2021foveated}. Ray tracing engines such as TACC Galaxy~\cite{abram2018galaxy} (\url{https://github.com/tacc/galaxy}) offer distributed asynchronous in situ visualization capabilities for regional-scale simulations. 

In this study, we develop an in situ visualization technique to simulate and render regional-scale landslides by coupling the CB-Geo MPM code for large-deformation modeling with distributed asynchronous ray tracing engine TACC Galaxy. We apply the developed in situ visualization technique in MPM to simulate and visualize the disastrous 2014 Oso landslide, also known as the SR 530 landslide, which occurred in the northwest Washington state in the US,  resulting in an 1100 m runout~\cite{yerro2019runout}. 

\begin{figure}[!htbp]
    \centering
    \includegraphics[width=\linewidth]{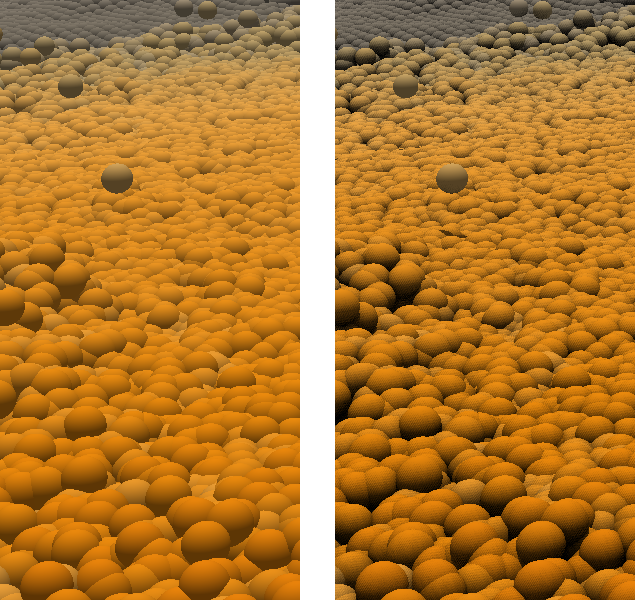}
    \caption{Detail of visualization without (L) and with (R) ambient occlusion and shadow global illumination.} 
    \label{fig:global_illum}
\end{figure}

\Cref{fig:oso3,fig:timing} show visualizations of the MPM simulation at different stages of the Oso landslide runout.   In this case, we use Galaxy to render the scene using spheres to represent the material points in the simulation, each with a fixed radius matching the original particle grid resolution and colored by the distance between each particle's current and original location (i.e., particle displacement).  In this study, we use Galaxy's global illumination model - shadows and ambient occlusion - to provide additional depth perception.  \Cref{fig:global_illum} shows a side-by-side comparison of detail of a timestep visualizations with and without global illumination.    For a discussion of the impact of these techniques on the ability to discern complex shapes, please see Grottel et. al \cite{Grottel12}.

\section{Related work}
\subsection{Visualization Systems}
Systems for visualizing large-scale scientific data have been available for decades.  High-performance graphics engines, add-ons to powerful workstations, first made interactive post-processing visualization of large datasets possible in the 1980s, leading to visualization systems like AVS, IBM Data Explorer, and Silicon Graphics IrisExplorer.  These systems and others forged a visualization paradigm in which non-sophisticated users, i.e., non-programmers, could use graphical user interfaces to develop specific visualizations of their data and provide real-time interactive data exploration.  

In the early days, such high-performance visualization engines were expensive and capable only of low-level rendering functions and were limited to dedicated visualization systems.  This separation of simulation and visualization led to a distinction between general-purpose computation systems for running simulations and graphics-capable systems for visualization.  Based on a \textit{post-processing} workflow paradigm, simulations would run first on general-purpose systems, then transferred to separate systems for visualization.  Post-processing remains the standard approach to visualizing large-scale data.

Today, simulations have outstripped the capabilities of even the most heavyweight conventional processors, leading to the development of today's supercomputers - \textit{clusters} of conventional processors (referred to as \textit{nodes}) with very high-speed intercommunications. Super computers require simulation applications, such as MPM, incorporate sophisticated algorithms to distribute the workload over many cooperating nodes. To keep up, visualization systems have also adopted a distributed-parallel architecture; popular general-purpose systems (notably ParaView and VisIt) have been adapted to distributed-memory parallel systems to provide visualization capabilities for simulations running on the largest super computing systems in the world.

\subsection{In situ Visualization}

Unfortunately, the I/O capabilities of such systems have not matched their computational capabilities, making it even more expensive to export data from the simulation system memory to secondary storage and subsequently read it back into visualization system memory.   Expensive I/O leads to a tension between the expense of saving time-step data and visualization requirements.
  
\textit{In situ} visualization is a broad approach to processing simulation data in real-time - that is, \textit{wall-clock} time, as the simulation is running.  Generally, the approach is to provide \textit{data extracts}, which are condensed representations of the data chosen for the explicit purpose of visualization and computed without writing data to external storage.  Since these extracts (often images) are vastly smaller than the raw simulation itself, it becomes possible to save them at a far higher temporal frequency than is practical for the raw data, resulting in substantial gains in both efficiency and accuracy.  In situ visualization allows simulations to export complete datasets only at the temporal frequency necessary for economic checkpoint/restart. 
 
Many approaches to in situ visualization exist.  Childs et al.\cite{Childs:IJHPCA} provide an exhaustive taxonomy of approaches, differentiating along six axes: the type of integration between simulation and visualization components,  how the components are situated on computing resources, how the visualization components access simulation data, timing, control, and output type.  For a complete discussion, please refer to the paper~\cite{Childs:IJHPCA}.  The paper also offers a survey of existing in situ systems.  Both ParaView and VisIt provide in situ capabilities; each provides an interface (Catalyst and libsim) with which data is transferred from the simulation code to co-resident visualization components; the simulation code then transfers control to the visualization components, resuming only when the visualization components return.
 
\subsection{Rendering Techniques}
At run-time, visualization consists of two phases: converting raw data to renderable primitives (generally triangles) and transferring data to a rendering subsystem that produces pixels in an image buffer.  Traditionally, this rendering process uses the \textit{rasterization} and Z-buffer algorithms to generate pixels. This approach, implemented in hardware, considers each primitive independently, and iterates over all the pixels to determine which pixels are affected by the primitive and evaluates how far each affected pixel is from the viewpoint. If a given affected pixel is the closest so far (by comparing depth with the current value in the Z-buffer), it is shaded and saved to the image buffer, and its depth is saved to the Z-buffer.  Since this algorithm touches all the primitives, the cost is O(n), where n is the number of primitives.

More recently, this approach is being replaced by \textit{ray-tracing}.  In this algorithm, the rendering loop is across all pixels in the image, asking which primitive is visible at the pixel.  This algorithm relies on an O(log n) search for each pixel and, on today's hardware (both general-purpose multi-core vector CPUs and graphics cards), is competitive in speed with rasterization and provides additional capabilities.  Ray tracing provides a simple approach to realistic lighting effects - notably shadowing and ambient occlusion - that are hard to implement in rasterization algorithms.  Both VisIt and ParaView optionally replace their traditional scan-conversion rendering with ray tracing, utilizing Intel's CPU-based OSPRay and NVIDIA's GPU-based Optix.  

Another advantage of the ray tracing algorithm is that, in some critical cases, it avoids the necessity of reducing data to low-level geometric primitives.  Critical to the current work is that the rendering of spheres in a rasterization algorithm requires that each sphere be represented as a sufficiently large set of triangles enough to give the appearance of a smooth sphere.   When there is a large number of spheres to render, rasterization (which considers every triangle) becomes very expensive.  Instead, using ray tracing, we can represent each sphere by its center and radius; rather than exhaustively intersect a ray against each triangle of a sphere's tessellated representation, we can determine the distance between the ray and the center and ask whether it is less than the radius.  This feature is apparent in ParaView; while the traditional rasterization back end required a filter to create spheres of each point, the points can be rendered directly using the OSPRay backend with substantially improved performance.

\subsection{Parallel Rendering}
When visualizing on multiple nodes in parallel, the resulting image may contain imagery computed on any participating node. To get a correct image, visualization systems generally use one of two algorithms, described in Molnar et. al ~\cite{Molnar94} as \textit{sort-first\textit} and \textit{sort-last} rendering. The sort-first algorithm redistributes the data to provide all the information necessary to render a screen patch (possibly the entire image).  In the sort-last algorithm, each node renders its portion of the data independently and then uses a distributed-parallel depth-based compositing algorithm to reconcile these partial results into a final image. Each has its advantages: if the amount of data required to render the image is small, it can be gathered cheaply and then rendered interactively with no further data movement.  On the other hand, when the data is large, the gather is expensive (and potentially exceeds the memory available on the rendering node(s)), and the post-rendering compositing step is preferable.

When advanced global lighting is required, we note that sort-last rendering will not work.  Without shadows, the correct shading for a surface is determined independently of other surfaces and thus can be performed locally before compositing.  However, when shadows are required, it is possible that a surface can be shadowed by another surface \textit{that is not present on current node}.  Thus post-rendering compositing cannot create a correct image, leaving gathering the data as the alternative.

Galaxy, developed at TACC, instead relies on \textit{spatial partitioning of the data} to implement Molnar's \textit{sort-middle} algorithm.  Spatial partitioning - generally a natural data decomposition in HPC simulation - ensures that all the data that may be relevant in region of space is resident on the process responsible for that region.  Galaxy's cooperating processes can therefore trace rays \textit{within their partition space} independently of data in other spaces.  Within the partition, rays are either retired (by reaching full opacity in volume rendering, or by encountering surfaces), or they reach the partition boundary.   When a boundary is encountered, the rays are transmitted to the process responsible for the partition across the boundary.  Thus the lighting of a surface hit point within a partition can be resolved by a secondary ray that first determines whether the light is occluded by local data, and if not, by data resident in other processes on the way to the light source.

Note that Galaxy is actually a framework for the management and exchange of rays between processes; it does not actually perform ray/surface or ray/volume intersections itself.   Rather, it relies on OSPRay (i.e., OSPRay's Embree and OpenVKL underpinnings) to perform the \textit{local} ray tracing.  This could equally well be performed by Optix.   For more information on Galaxy, please see Abram et. al. \cite{abram2018galaxy}.

\subsection{Providing Interactive Visualization In Situ}
Of course, in situ visualization largely precludes exploratory visualization: the in situ analysis will only produce the anticipated results at set-up time.  With post-processing, the entire dataset is available when the user sits down in front of the visualization system; the user can adjust viewpoints and parameters interactively during their session, a powerful capability lost in in situ visualization. However, Ahrens et al. \cite{Ahrens2014cinema} observed that, for the cost of writing a single timestep of the simulation data, we could save many \textit{images} of the timestep.  Therefore, we can set up an in situ process that generates as many images of the data - varying viewpoints and parameters - as the user anticipates and wants. Offering in situ visualization capabilities incurs additional compute cost at run-time. However, post-processing visualization also has this addition compute cost, which is likely to be higher than in situ visualization due to increased storage and latency requirements. Ahrens et al. have developed a database format storing these images, and Cinema - a browser for the interactive exploration of such databases.  We can, for example, specify a range of viewpoints in angle and distance and generate an image for each timestep; the Cinema browser will then present an interactive window giving the appearance of being able to interact with the data in 4 dimensions in real-time.

\section{Study Detail}
The 3D Oso landslide region of interest spans $1500 \times 1500 \times 270$ m. This domain is discretized into 4.2 million material points and a background grid with a cell size of $8 \times 8 \times 8$ m. In most regions, we use two material points in each direction (8 material points per cell), while the landslide runout region has a higher resolution with four material points in each direction (64 points per cell). A Mohr-Coulomb material model with a Total Stress strength parameters is used to simulate the undrained runout evolution of the Oso landslide in MPM. 

In this work, we consider in situ visualization for MPM. In particular, we note that the data for visualization is simple: a large set of points (currently ~4.2M and potentially billions) with scalar attributes, scattered among a large number of parallel processes.  The objective is to generate potentially many images at an appropriate temporal frequency and minimize the need for MPM to dump timestep data.  We would like to do so with minimal impact on both the MPM codebase (minimizing additional code and dependencies) and the run-time, minimizing the impact on MPM's run-time memory footprint and performance.  Finally, we would like this interface to be intermittent, allowing us to connect a visualization to a running instance of MPM, enabling us to examine the current state of the simulation.

Ray tracing is a natural solution for rendering MPM particle data.  Rather than represent each particle as a set of primitive surface elements (polygons, necessary for traditional scan-conversion rendering), we can represent particle data far more compactly as a center point and radius.  

We interface Galaxy, our ray-tracing-based visualization system, and CB-Geo MPM to perform in situ visualization of the Oso landslide.  As described above, Galaxy is able to perform global illumination on our data \textit{without gathering all the data to each node}.  Galaxy has also been shown to scale to over a billion particles and demonstrated to scale to 128 nodes; see the \textit{Cosmic Web} test case in Abram et. al. \cite{abram2018galaxy}.  

Two concerns arose in developing the MPM-Galaxy interface.   First, we would like to balance the performance of the MPM simulation with the Galaxy visualization by varying the ratio of MPM processes to Galaxy processes.   Second, the spatial partitioning layout in MPM is not absolute; while each MPM process manages a relatively compact region of space, these regions overlap and do not form the clean Cartesian boundaries necessary for Galaxy.   To address these problems, we developed an asynchronous M to N data transfer scheme between M MPM processes and N Galaxy processes that incorporates a sorting step.   This is implemented by communicating an appropriate layout of partitions, given the number of Galaxy processes in play and the bounding box of the MPM computational space, to MPM at data transfer time.   Each MPM process then traverses its particle data, determines which Galaxy process's partition space the particle lies, and transfers them accordingly.  The M:N point-to-point interface between M MPM worker processes and N Galaxy processes offers better load balancing performance.  We can vary the number of processes for MPM and Galaxy so that they run at approximately the same pace.  When out of balance, we can
choose to drop frames of the visualization by having MPM
continue without pausing for Galaxy (if Galaxy is not ready), or by having MPM wait for Galaxy to be ready.

\subsection{The Interface}

We use a sockets-based interface for this preliminary work, enabling the two components to communicate via the IP protocol.  While this enables us to position the components anywhere with IP connectivity, by placing the two on the same supercomputer, we are able to use the full parallel bandwidth of the supercomputer's interconnect to communicate between MPM and Galaxy worker processes.

We use a single connection between the root processes of MPM and Galaxy to control the overall process. When Galaxy is launched, its root process creates a master server socket and waits for the simulation's root process to connect. Each worker process opens a server socket and waits for zero or more simulation workers to connect.

Periodically, the MPM root process (at its synchronization point) attempts to connect to this server via a socket. A small amount of control information is passed when the connection is established, including a list of host/port pairs for listening Galaxy worker processes and the number of MPM worker processes transferring data. This lets each MPM worker know where to send its data, and each Galaxy worker knows how many connections to expect. Data transfer occurs in parallel, and MPM continues when the transfers are complete.

\subsection{Data}

We have implemented an ad-hoc data structure for passing data from MPM to Galaxy for this preliminary work.  This format consists of a small header containing the name of the variable being transferred and the number of points/data values, followed by a block of binary floats comprising the points, followed by the binary float scalar data values for each point.

\subsection{Galaxy Side}

Once data has arrived on the Galaxy side, it is re-distributed into a Galaxy-compliant partitioning of the computational space.  As mentioned above, Galaxy relies on a regular subdivision of space - which is not the case for data acquired from MPM, particularly when multiple MPM workers can send data to each Galaxy node.  Since each Galaxy worker knows the overall partitioning, this reshuffle is performed using point-to-point messaging via MPI.  Each worker reads and writes approximately 1/N of the data after examining 1/N of the data.  We note that while this repartitioning accrue cost, this cost is amortized across all visualization time steps~\cite{abram2018galaxy}.

\subsection{Specifying the Visualization}

Galaxy uses a JSON text file (a \textit{state} file) to specify a visualization.  The JSON consists of several sections describing: the data objects, the distinct visualizations, listing the data objects and operators to be applied to the lighting environment, a list of cameras, and renderer parameters.  Names in the data object section correspond to the variables in MPM. At setup time, Galaxy creates empty distributed containers for each named object.  At run time, as the material points traverse the background grid, the reshuffling results from the MPM simulations (if any) will replace the contents of the distributed datasets.  

From the state file, Galaxy generates a \textit{RenderingSet}, consisting of a list of individual renderings, one for each visualization/camera pair.  Executing the RenderingSet causes the renderings to be initiated simultaneously, and individual image files are generated.

Galaxy has a state-file generator program for Cinema.  A \textit{.cinema} file is similar to that specifies \textit{ranges} of parameters, \textit{cinema2state} generates a single state file that contains a list of the visualizations described in the input file with every combination of parameter settings, and a list of cameras with all the combinations of viewpoint parameters.  Galaxy will generate a complete Cinema database with an interactor for each varying parameter and time.

\section{Results}

\begin{figure*}[!tbp]
    \begin{subfigure}[b]{0.495\textwidth}
        \includegraphics[width=\textwidth]{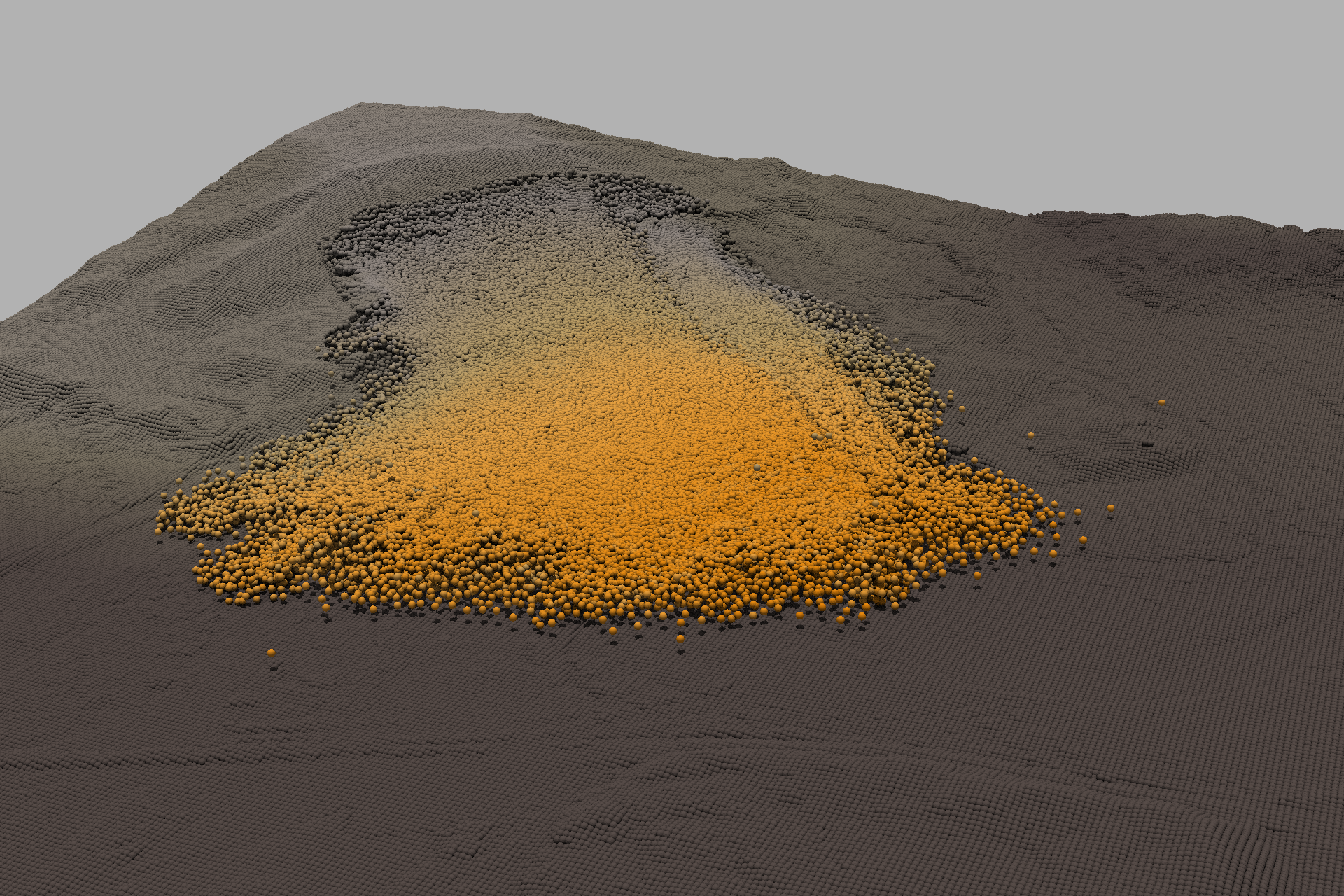}
        \caption{runout after 50 s}
        \label{fig:oso-50} 
    \end{subfigure}
    \begin{subfigure}[b]{0.495\textwidth}
        \includegraphics[width=\textwidth]{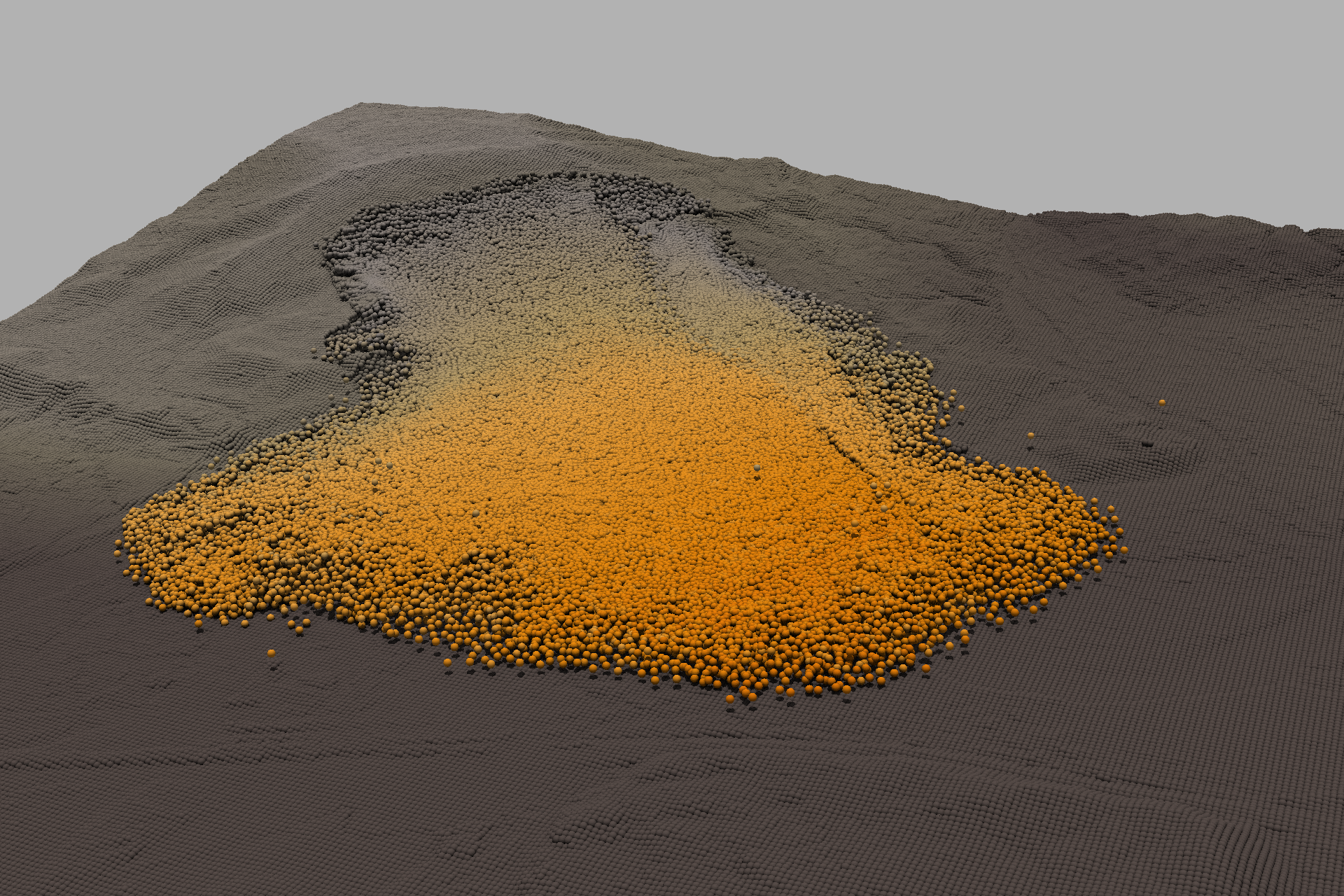}
        \caption{runout after 74 s}
        \label{fig:oso-74} 
    \end{subfigure}
    \caption{In situ rendering of MPM simulation with Galaxy. The color gradient shows the amount of particle displacements from their original position.}
    \label{fig:timing}
\end{figure*}

Given a checkpoint restart file at step 50,000 (actual runout time of 50 s) of the MPM Oso simulation, we ran two versions of the next 25,000 steps to evaluate the cost of in situ visualization on the simulation.   Using four nodes of Stampede2 at TACC, each running four MPM worker processes, we first ran 15,000 steps without in situ visualization but saving checkpoint files every 5,000 steps, resulting in five time step datasets over this interval. In this case, the wall-clock time required to run each 5,000 steps - as determined by timestamps on output files - was approximately 2 hours, 23 minutes.   Figures \ref{fig:oso-50} and \ref{fig:oso-74} visualize the state of this run at the beginning and end of the simulation interval, using a color-scale to show the distance each particle has traveled from its original location.  Clearly, the simulation has made substantial progress in this time interval.

We then re-ran the same interval of the simulation with the same checkpoint dumps, but now using the MPM/Galaxy interface to transfer data from MPM to Galaxy every 100 steps (0.1 s of real-time runout, where a difference in the runout can be perceived).  In situ visualization with Galaxy/MPM integration offers insights into the dynamics of very fast runouts, such as the Oso landslide (total duration of runout is 120s) with a very fine temporal resolution, which are otherwise infeasible using traditional post-processing visualization. We found that Galaxy was substantially faster than MPM, even when running on a single node.  To make full use of the I/O bandwith available on that node, we use a M:N interface of 4:1 with four MPM worker nodes to one Galaxy worker nodes.  With this setup, Galaxy required approximately 6.5 seconds for each timestep to receive the data, repartition it, and render the visualizations - far less than the three minutes required by MPM to run through 100 steps.   In this second case with in situ visualization, the wall-clock time needed by MPM to run through 5,000 steps was approximately 2 hours, 26 minutes, or 2 percent more than the first, non-visualized case.  Galaxy renderings of the Oso landslide during the early and late stages of the runout are shown in~\cref{fig:oso3,fig:timing}.

Galaxy supports including additional views at runtime as we observe interesting dynamics for visualization as the simulation progresses.  Galaxy initially generated two images per timestep and later three, when we observed the runout spreading outside our initial camera views.

\section{Conclusion}

In our study with 4.7 million material points, the time cost of in situ data transfer from simulation to a visualization application located on a separate partition of the same supercomputer is negligible.   A many-to-many communications strategy using point-to-point communications enables us to use the full cross-sectional internal bandwidth of the supercomputer interconnect to move data from a simulation to an external concurrently running analysis component with minimal impact on the simulation.  
Further, the time required to generate the three simple views of the data for every 100 timesteps is much less than the cost of simulating 100 timesteps - though clearly this depends on both the simulation and the visualization.   This leads to several options.  We could reduce the resources allocated to visualization for better balance of \textit{this} workflow (though not in this case, since we use a single node for visualization).  Alternatively, we could increase the work necessary on the visualization side, perhaps by generating \textit{many} views of each timestep, enabling Cinema-based, real-time interaction with visualization \cite{Ahrens2014cinema}.   Finally, we could increase the frequency of timesteps.    In any event, our architecture enables these factors to be considered independently, by varying the allocation of resources to the in situ analysis.

Clearly, the test example we have worked with - the Oso landslide at a relatively coarse resolution, does not tax our system.   However, as observed above, regional-scale simulation of landslides resulting in kilometer-scale runout will require billions of particles rather than millions.   We believe our approach - notably using a scalable distributed visualization component (Galaxy, \cite{abram2018galaxy}) and a many-to-many, point-to-point, in situ interface that leverages the parallel internal bandwidth of the supercomputer, provides a means for effectively visualizing full-scale simulations.

\appendices

\section*{Acknowledgment}
The simulations and in situ visualization of the Oso landslide were conducted on the Texas Advanced Computing Center’s (TACC’s) cluster Stampede2.  Krishna Kumar would like to thank the U.S. National Science Foundation grant CMMI-2022469. Greg Abram would like to thank the U.S. Department of Energy grant DE-SC0012513. Andrew Solis and Greg Abram would like to thank the TACC Intel Graphics and Visualization Institute of eXcellence. However, any opinions, findings and conclusions or recommendations expressed in this material are those of the authors and do not necessarily reflect the views of the National Science Foundation.

\ifCLASSOPTIONcaptionsoff
  \newpage
\fi



\bibliographystyle{IEEEtran}
\bibliography{IEEEabrv,references.bib}

\begin{thebibliography}{10}
\providecommand{\url}[1]{#1}
\csname url@samestyle\endcsname
\providecommand{\newblock}{\relax}
\providecommand{\bibinfo}[2]{#2}
\providecommand{\BIBentrySTDinterwordspacing}{\spaceskip=0pt\relax}
\providecommand{\BIBentryALTinterwordstretchfactor}{4}
\providecommand{\BIBentryALTinterwordspacing}{\spaceskip=\fontdimen2\font plus
\BIBentryALTinterwordstretchfactor\fontdimen3\font minus
  \fontdimen4\font\relax}
\providecommand{\BIBforeignlanguage}[2]{{%
\expandafter\ifx\csname l@#1\endcsname\relax
\typeout{** WARNING: IEEEtran.bst: No hyphenation pattern has been}%
\typeout{** loaded for the language `#1'. Using the pattern for}%
\typeout{** the default language instead.}%
\else
\language=\csname l@#1\endcsname
\fi
#2}}
\providecommand{\BIBdecl}{\relax}
\BIBdecl

\bibitem{kim2020public}
\BIBentryALTinterwordspacing
D.~K.~D. Kim and T.~P. Madison, ``Public risk perception attitude and
  information-seeking efficacy on floods: a formative study for disaster
  preparation campaigns and policies,'' \emph{International Journal of Disaster
  Risk Science}, vol.~11, no.~5, pp. 592--601, 2020. [Online]. Available:
  \url{https://doi.org/10.1007/s13753-020-00307-5}
\BIBentrySTDinterwordspacing

\bibitem{kahneman1983cost}
\BIBentryALTinterwordspacing
D.~Kahneman, A.~Treisman, and J.~Burkell, ``The cost of visual filtering.''
  \emph{Journal of Experimental Psychology: Human Perception and Performance},
  vol.~9, no.~4, p. 510, 1983. [Online]. Available:
  \url{https://doi.org/10.1037/0096-1523.9.4.510}
\BIBentrySTDinterwordspacing

\bibitem{soga2016trends}
\BIBentryALTinterwordspacing
K.~Soga, E.~Alonso, A.~Yerro, K.~Kumar, and S.~Bandara, ``Trends in
  large-deformation analysis of landslide mass movements with particular
  emphasis on the material point method,'' \emph{G{\'e}otechnique}, vol.~66,
  no.~3, pp. 248--273, 2016. [Online]. Available:
  \url{https://doi.org/10.1680/jgeot.15.LM.005}
\BIBentrySTDinterwordspacing

\bibitem{kumar2019scalable}
\BIBentryALTinterwordspacing
K.~Kumar, J.~Salmond, S.~Kularathna, C.~Wilkes, E.~Tjung, G.~Biscontin, and
  K.~Soga, ``Scalable and modular material point method for large-scale
  simulations,'' in \emph{2nd International Conference on the Material Point
  Method for Modelling Soil–Water–Structure Interaction}, 2019. [Online].
  Available: \url{https://doi.org/10.31224/osf.io/e24rb}
\BIBentrySTDinterwordspacing

\bibitem{byna2020exahdf5}
\BIBentryALTinterwordspacing
S.~Byna, M.~S. Breitenfeld, B.~Dong, Q.~Koziol, E.~Pourmal, D.~Robinson,
  J.~Soumagne, H.~Tang, V.~Vishwanath, and R.~Warren, ``Exahdf5: delivering
  efficient parallel i/o on exascale computing systems,'' \emph{Journal of
  Computer Science and Technology}, vol.~35, no.~1, pp. 145--160, 2020.
  [Online]. Available: \url{https://doi.org/10.1007/s11390-020-9822-9}
\BIBentrySTDinterwordspacing

\bibitem{yang2021foveated}
\BIBentryALTinterwordspacing
Q.~Yang, Z.~Chen, Y.~Liu, G.~Xing, and Y.~Zhang, ``Foveated light culling,''
  \emph{Computers \& Graphics}, vol.~97, pp. 200--207, 2021. [Online].
  Available: \url{https://doi.org/10.1016/j.cag.2021.04.021}
\BIBentrySTDinterwordspacing

\bibitem{abram2018galaxy}
\BIBentryALTinterwordspacing
G.~Abram, P.~Navr{\'a}til, P.~Grossett, D.~Rogers, and J.~Ahrens, ``Galaxy:
  Asynchronous ray tracing for large high-fidelity visualization,'' in
  \emph{2018 IEEE 8th Symposium on Large Data Analysis and Visualization
  (LDAV)}.\hskip 1em plus 0.5em minus 0.4em\relax IEEE, 2018, pp. 72--76.
  [Online]. Available: \url{http://doi.org/10.1109/LDAV.2018.8739241}
\BIBentrySTDinterwordspacing

\bibitem{yerro2019runout}
\BIBentryALTinterwordspacing
A.~Yerro, K.~Soga, and J.~Bray, ``Runout evaluation of oso landslide with the
  material point method,'' \emph{Canadian Geotechnical Journal}, vol.~56,
  no.~9, pp. 1304--1317, 2019. [Online]. Available:
  \url{https://doi.org/10.1139/cgj-2017-0630}
\BIBentrySTDinterwordspacing

\bibitem{Grottel12}
\BIBentryALTinterwordspacing
S.~Grottel, M.~Krone, K.~Scharnowski, and T.~Ertl, ``Object-space ambient
  occlusion for molecular dynamics,'' \emph{2012 IEEE Pacific Visualization
  Symposium}, 2012. [Online]. Available:
  \url{https://doi.org/10.1109/PACIFICVIS.2012.6183593}
\BIBentrySTDinterwordspacing

\bibitem{Childs:IJHPCA}
\BIBentryALTinterwordspacing
H.~Childs, S.~D. Ahern, J.~Ahrens, A.~C. Bauer, J.~Bennett, E.~W. Bethel, P.-T.
  Bremer, E.~Brugger, J.~Cottam, M.~Dorier, S.~Dutta, J.~M. Favre, T.~Fogal,
  S.~Frey, C.~Garth, B.~Geveci, W.~F. Godoy, C.~D. Hansen, C.~Harrison,
  B.~Hentschel, J.~Insley, C.~R. Johnson, S.~Klasky, A.~Knoll, J.~Kress,
  M.~Larsen, J.~Lofstead, K.-L. Ma, P.~Malakar, J.~Meredith, K.~Moreland,
  P.~Navr\'{a}til, P.~O'Leary, M.~Parashar, V.~Pascucci, J.~Patchett,
  T.~Peterka, S.~Petruzza, N.~Podhorszki, D.~Pugmire, M.~Rasquin, S.~Rizzi,
  D.~H. Rogers, S.~Sane, F.~Sauer, R.~Sisneros, H.-W. Shen, W.~Usher,
  R.~Vickery, V.~Vishwanath, I.~Wald, R.~Wang, G.~H. Weber, B.~Whitlock,
  M.~Wolf, H.~Yu, and S.~B. Ziegeler, ``{A Terminology for In Situ
  Visualization and Analysis Systems},'' \emph{International Journal of High
  Performance Computing Applications (IJHPCA)}, vol.~34, no.~6, pp. 676--691,
  Nov. 2020. [Online]. Available:
  \url{https://doi.org/10.1177/1094342020935991}
\BIBentrySTDinterwordspacing

\bibitem{Molnar94}
\BIBentryALTinterwordspacing
S.~Molnar, M.~Cox, D.~Ellsworth, and H.~Fuchs, ``A sorting classification of
  parallel rendering,'' \emph{IEEE Computer Graphics and Applications},
  vol.~14, no.~4, pp. 23--32, 1994. [Online]. Available:
  \url{https://doi.org/10.1109/38.291528}
\BIBentrySTDinterwordspacing

\bibitem{Ahrens2014cinema}
\BIBentryALTinterwordspacing
J.~Ahrens, S.~Jourdain, P.~O'Leary, J.~Patchett, D.~H. Rogers, and M.~Petersen,
  ``An image-based approach to extreme scale in situ visualization and
  analysis,'' in \emph{Proceedings of the International Conference for High
  Performance Computing, Networking, Storage and Analysis}, ser. SC '14.\hskip
  1em plus 0.5em minus 0.4em\relax Piscataway, NJ, USA: IEEE Press, 2
  -1-4799-5500-8, pp. 424--434. [Online]. Available:
  \url{https://doi.org/10.1109/SC.2014.40}
\BIBentrySTDinterwordspacing

\end{thebibliography}
%\bibliography{references.bib}


% \end{IEEEbiography}

%\vfill\null
%\columnbreak
\vspace{-0.5cm}
\begin{IEEEbiographynophoto}{Greg Abram}
is a visualization researcher in the at the Texas Advanced Computing Center, University of Texas at Austin, USA, 78758. His research interests focus on parallel visualization systems.   Dr. Abram received a Ph.D. in Computer Science from the University of North Carolina at Chapel Hill. Contact him at gda@tacc.utexas.edu.
\end{IEEEbiographynophoto}
\vspace{-0.5cm}
\begin{IEEEbiographynophoto}{Andrew Jay Solis}
is a visualization researcher in the at the Texas Advanced Computing Center, University of Texas at Austin, USA, 78758. His research interests include extended realities, HCI, scientific computation, and parallel computing. Solis received a B.S. in Computer Science from the University of Texas at Austin. Contact them at asolis@tacc.utexas.edu.
\end{IEEEbiographynophoto}
\vspace{-0.5cm}
\begin{IEEEbiographynophoto}{Yong Liang}
is a postdoctoral researcher in the Civil and Environmental Engineering, University of California at Berkeley, USA, 94720. His research interests include fracture modeling and material point method. Yong received his Ph.D. from Tsinghua University, China. Contact them at yliang\_sn@berkeley.edu.
\end{IEEEbiographynophoto}
\vspace{-0.5cm}
\begin{IEEEbiographynophoto}{Krishna Kumar}
is an Assistant Professor in the Civil, Architectural and Environmental Engineering, University of Texas at Austin, USA, 78751. His research interests include high-performance computing, numerical modeling, and explainable-AI. Krishna received his Ph.D. in Engineering from the University of Cambridge, UK. Contact them at krishnak@utexas.edu.
\end{IEEEbiographynophoto}
\vfill

\end{document}